\pdfoutput=1
\documentclass[a4paper]{article}
\usepackage{cite}
\usepackage{INTERSPEECH2020}
\usepackage{todonotes}
\title{Learning Joint Articulatory-Acoustic Representations with Normalizing Flows}
\name{Pramit Saha$^1$ and Sidney Fels$^2$}

\address{
  $^1$Department of Electrical and Computer Engineering, University of British Columbia, Canada}
\email{pramit@ece.ubc.ca, ssfels@ece.ubc.ca}

\begin{document}

\maketitle
\begin{abstract}

The articulatory geometric configurations of the vocal tract and the acoustic properties of the resultant speech sound are considered to have a strong causal relationship. This paper aims at finding a joint latent representation between the articulatory and acoustic domain for vowel sounds via invertible neural network models, while simultaneously preserving the respective domain-specific features. Our model utilizes a convolutional autoencoder architecture and normalizing flow-based models to allow both forward and inverse mappings in a semi-supervised manner, between the mid-sagittal vocal tract geometry of a two degrees-of-freedom articulatory synthesizer with 1D acoustic wave model and the Mel-spectrogram representation of the synthesized speech sounds. Our approach achieves satisfactory performance in achieving both articulatory-to-acoustic as well as acoustic-to-articulatory mapping, thereby demonstrating our success in achieving a joint encoding of both the domains.

\end{abstract}


\section{Introduction}
The vocal tract (VT) geometry \cite{story1996vocal,saha2018towards} and its dynamic changes play a crucial role in generating distinguishable speech sounds, via modulation of airflow and creation of various resonant cavities inside the tract. Therefore the acoustic signal contains abundant information that can be extracted for better understanding the underlying upper airway geometries involved in speech production mechanism. Determining the relationship between the articulatory gestures and acoustic parameters has been a long-standing issue in related research areas  \cite{mitra2017joint,hueber2011statistical,saha2018towards,hiroya2004estimation,hodgen2001stochastic, saha2020ultra2speech}. 

The aforementioned problem is generally classified into two broad types : (a) articulatory-to-acoustic mapping, also known as forward mapping problem and (b) acoustic-to-articulatory inversion or speech inversion, also known as inverse mapping problem . The former deals with the production of audio speech signals from vocal tract, thereby modeling variations in acoustic space with variations in vocal tract shapes, while, the latter one encompasses the recovery of vocal tract configurations responsible for production of given speech signals. 
The forward mapping, \textit{i.e.}, estimating acoustic response to articulatory behaviour is of utmost importance in the development of articulatory speech synthesizers and other silent speech interfaces as well as in detailed study of speech production and articulatory phonetics. On the other hand, applications of the inverse mapping, \textit{i.e.}, inferring articulatory information from speech acoustics include estimation of vocal tract parameters for efficient speech coding, enhanced speech recognition systems and for developing visual articulatory feedback systems. The related works mostly address either of these two problems independently and as such, there is a lack of unified end-to-end forward and inverse mapping approach that can reversibly map the vocal tract shapes and corresponding speech sounds. This is because it is incredibly challenging to accurately determine a joint distribution of the articulatory and acoustic domains, both having complex generative processes involving a series of motor control and estimation tasks, biomechanical mechanisms and aero-dynamic flow - some being shared across both generations while some being specifically important to one of them.

In order to address this issue, we employ a semi-supervised, invertible, bijective cross-domain mapping between vocal tract geometries and the acoustic outputs, leveraging a pair of deep convolutional autoencoders and normalizing flow based probability density estimation technique. In this paper, we particularly consider the mid-sagittal vocal tract configurations and synthesized vowel sounds, simulated in the online articulatory speech synthesizer application named Pink Trombone \cite{pt}, as our input-output space. Our approach involves a separated yet shared encoding of the images, capturing diverse vocal tract shape, as well as the mel-spectrograms, possessing the acoustic information pertaining to the resultant speech signals, in an unsupervised manner. The double autoencoders are simultaneously aligned in a supervised fashion by stacking a chain of invertible bijective transformation functions between the bottleneck feature distributions. The core idea is to constrain the latent representations of both the domains to have some domain-specific features pertaining to self-reconstruction as well as a joint feature space that encodes the mutual characteristics for enabling cross domain VT geometry-to-speech and speech-to-VT geometry synthesis. Furthermore, the domain-specific latent codes is kept conditional on the shared cross-domain latent space by enforcing a normalizing flow \cite{kingma2018glow} based conditional prior in the articulatory-acoustic latent representation. In the next section, we will lay the foundation of our approach and present a systematic study on the variational model employed to achieve the target mapping.
\section{Proposed Mapping Strategy}

\subsection{Problem formulation and overview}
In order to investigate the joint distribution of vocal tract shapes and acoustics $p(x_{g},x_{s})$ which follow the generative processes $p_{g}(x_{g})$ and $p_{s}(x_{s})$ respectively, we define a common latent variable $z$ such that the marginal likelihood $p(x_{g},x_{s})=\int p(x_{g},x_{s},z)~ dz$, where the joint probability distribution $p(x_{g},x_{s},z)=p(x_{g},x_{s}|z)p(z)$. The likelihood $p(x_{g},x_{s}|z)$ indicates the probability distribution over the observed variables in articulatory and acoustic space, given the latent representation $z$. The standard practise is to compute the likelihood using the posterior distribution $p(z|x_{g},x_{s})$ via Bayes' rule. However computing the posterior distribution is intractable in general as there exists no closed form solution. Alternatively, a variational distribution $\Psi(z|x_{g},x_{s})$ is used to approximate the posteriori by optimizing the evidence lower bound (ELBO) \cite{kingma2013auto}. Therefore in our case, the maximization of the likelihood of $p(x_{g},x_{s})$ can be achieved by involving a posterior encoding distribution $\Psi_{\phi}(z|x_{g},x_{s})$ parameterized by $\phi$. 
As such, our objective boils down to learning the variational posterior distribution $\Psi_{\phi}(z_{\widehat{gs}}|x_{g},x_{s})$ related to the shared latent space $(z_{\widehat{gs}})$ between the VT geometry $(x_{g})$ and the acoustic representation $(x_{s})$. Considering that the shared latent variable is capable of encoding joint articulatory and acoustic information, this implies, for a given pair of articulatory-acoustic data sample $(x_{g},x_{s})$, the learnt posterior distributions, $\Psi_{\phi}(z_{\widehat{gs}}|x_{g},x_{s})$ $\equiv$  $\Psi_{\phi}(z_{\widehat{gs}}|x_{g})$ $\equiv$ $\Psi_{\phi}(z_{\widehat{gs}}|x_{s})$. This can be ensured by enforcing the encoders in both the domains to generate same latent information. However, since the data distribution of articulatory and acoustic domains follow distinct underlying generative models as discussed earlier, it is not admissible to try to enforce exactly same latent variable representation by removing individual domain-specific information from the acoustic or articulatory space. For the same reason, minimizing the mean squared error between the encoder output features for enhancing shared information is not ideal. 

To this end, the shared information encoding is modified in two ways. The first is to partition the encodings of both the articulatory and acoustic domains into two parts: one that contains sole articulatory or acoustic information ($z_{g~\setminus~ \widehat{gs}}$ or $z_{s~\setminus~ \widehat{gs}}$) and the other which has the joint or shared information $(z_{\widehat{gs}})$. Therefore, our model consists of two domain-specific encoders : one for encoding the vocal tract geometry from the input image that learns an articulation-related posterior distribution $\Psi_{\gamma}(z_{g ~\setminus~ \widehat{gs}}|x_{g},x_{s})$ parameterized by $\gamma$ and the other for encoding the acoustic information from the mel spectrograms, that learns the latent posterior distribution of acoustic domain $\Psi_{\alpha}(z_{s ~\setminus~ \widehat{gs}}|x_{g},x_{s})$ parameterized by $\alpha$. 
The second modification is that, instead of constraining the encoders to learn the exact same shared latent encoding dimensions, we respect the constraints specific to articulation or acoustics and alternatively learn an invertible bijective mapping $\Omega_{\omega_{\widehat{gs}}} \colon \mathbb{R}^{d_{\widehat{gs}}} \xrightarrow{}\mathbb{R}^{d_{\widehat{gs}}}$ between the shared representation of the articulatory and acoustic space. This invertible mapping performs transformation of the ${d_{\widehat{gs}}}$ dimensional latent vector ${z_{\widehat{gs}}}$ between the domains $g$ and $s$.

\subsection{Self-attention based Convolutional Autoencoder}
The input-output space of our problem being artificial vocal tract images and mel-spectrograms, both are in the image representation. And our target is to encode the pair of image data into 2 sets of effective latent vectors or bottleneck features which best represent the respective domains and contain maximum relevant information required for their individual reconstruction. Convolutional architecture is a natural choice for the autoencoder network in this case as the convolutional autoencoders preserve the spatial information of the input image data, by incorporation of convolutional filter kernels in the network \cite{masci2011stacked}. Additionally, the self-attention mechanism of \cite{wang2018non,zhang2018self,vaswani2017attention} is also utilized in the encoder-decoder architecture as shown in Fig. 1, leveraging its capability of modeling non-local relationships between widely separated spatial regions - an equivalent of long-range dependency in images. 
\begin{figure}
	\centering
    \includegraphics[scale=0.25]{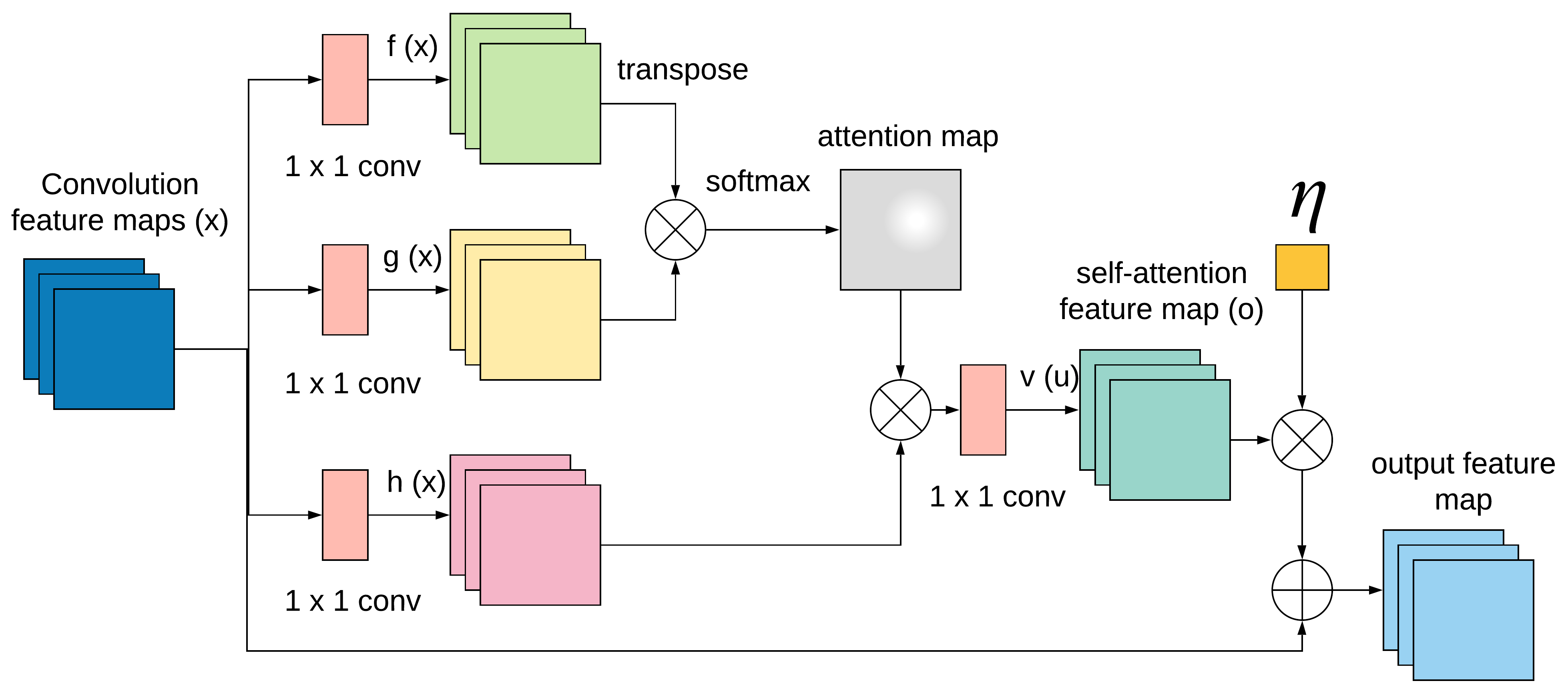}
    \caption{The self-attention module in convolutional autoencoder architecture}

    \vspace{-7pt}
\end{figure}
An attention map $\beta$ is generated by first transforming the feature set from the previous layer into two parallel layers $f(x)$ and $g(x)$ followed by exponentiating the product of these two feature sets and normalizing it as shown below:
\begin{equation}
 f(x)=W_{f}x,~~ x \in \mathbf{R}^{ C \times N},~ W_{f} \in \mathbf{R}^{ C \times C}
 \end{equation}
 \begin{equation}
g(x)=W_{g}x, ~W_{g} \in \mathbf{R}^{C \times C}
 \end{equation}
  \begin{equation}
\beta_{j,i}=\frac{f(x_{i})^{T}g(x_{j})}{\sum_{i=1}^{N} f(x_{i})^{T}g(x_{j})},~ \beta \in \mathbf{R}^{ N \times N}
 \end{equation}
 where $\beta_{j,i}$ denotes the impact of $i^{th}$ location  while rendering $j^{th}$ location, \textit{i.e.}, the extent to which the network attends to $i^{th}$ location  while synthesizing $j^{th}$ location. 
 
 Next, the previous layer features are again transformed to another feature set $h(x)$ and multiplied with the computed attention map $\beta$ to generate the self attention map output $o$.
 \begin{equation}
h(x)=W_{h}x, ~W_{h} \in \mathbf{R}^{C \times C}
 \end{equation}
  \begin{equation}
 v(u_{i})=W_{v}u_{i}
  \end{equation}
  \begin{equation}
o_{j}=v \left(\sum_{i=1}^{N}\beta_{j,i}~ h(x_{i})\right),  ~~o \in \mathbf{R}^{C \times N}
 \end{equation}
 $W_{f}$, $W_{g}$, $W_{h}$ and $W_{v}$ are learned weight matrices, implemented as $1\times1$ convolution operation. The final layer of the self-attention convolution layer is represented as the addition of a weighted self-attention mask (with the learnable scalar weight,  $\eta$) to the previous layer feature. 
 \begin{equation}
 y_{j}=\eta ~o_{j} + x_{j}, ~~ y\in \mathbf{R}^{C \times N}
 \end{equation}
 $\eta$ is initialized as 0 to let the model explore local spatial information before starting to capture non-local features via self-attention based refinement. 
 

Let the encoder networks corresponding to the vocal tract geometry $x_{g}$ of dimensions $d_{g}$ and the acoustic representation $x_{s}$ of dimensions $d_{s}$ be denoted as $\mathcal{G}_{\mathcal{E}_{g}}$ and $\mathcal{S}_{\mathcal{E}_{s}}$ with parameters ${\mathcal{E}_{g}}$ and ${\mathcal{E}_{s}}$ respectively, such that $\mathcal{G}_{\mathcal{E}_{g}} \colon (x_{g})_{d_{g}} \xrightarrow{}(z_{g})_{d^{l}_{g}}$ with ${\mathcal{E}_{g}}=\{\beta, \gamma\}$ and $\mathcal{S}_{\mathcal{E}_{s}} \colon (x_{s})_{d_{s}} \xrightarrow{}(z_{s})_{d^{l}_{s}}$ with ${\mathcal{E}_{s}}=\{\alpha, \gamma\}$, where $d^{l}_{g}$ and $d^{l}_{s}$ are the latent dimensions of articulatory and acoustic domains. Similarly, let the decoder networks corresponding to the vocal tract geometry and the acoustic representation be denoted as $\mathcal{G}_{\mathcal{D}_{g}}$ and $\mathcal{S}_{\mathcal{D}_{s}}$ with parameters ${\mathcal{D}_{g}}$ and ${\mathcal{D}_{s}}$ respectively, such that $\mathcal{G}_{\mathcal{D}_{g}} \colon (z_{g})_{d^{l}_{g}} \xrightarrow{}(x_{g})_{d_{g}}$ and $\mathcal{S}_{\mathcal{D}_{s}} \colon (z_{s})_{d^{l}_{s}} \xrightarrow{}(x_{s})_{d_{s}}$. Further, let the decoded vocal tract geometry and acoustic representation outputs be denoted as $\tilde x_{g}$ and $\tilde x_{s}$ respectively, then, $\tilde x_{g}=\mathcal{G}_{\mathcal{D}_{g}}(\mathcal{G}_{\mathcal{E}_{g}}(x_{g}))$ and $\tilde x_{s}=\mathcal{S}_{\mathcal{D}_{s}}(\mathcal{S}_{\mathcal{E}_{s}}(x_{s}))$. For vocal tract geometry image, the reconstruction loss between input image $(x_{g})$ and reconstructed image $(\tilde x_{g})$ from the image decoder is computed as $\mathcal{L}^{rec}_{g}(x_{g},\tilde x_{g})=\left\lVert x_{g} - \mathcal{G}_{\mathcal{D}_{g}}(\mathcal{G}_{\mathcal{E}_{g}}(x_{g}))\right\rVert$ where $\left\lVert .\right\rVert$ denotes $l_{2}$ norm. Similarly, for Mel-spectrogram image, the reconstruction loss between input Mel-spectrogram $(x_{s})$ and reconstructed Mel-spectrogram $(\tilde x_{s})$ from the spectrogram decoder is computed as $\mathcal{L}^{rec}_{s}(x_{s},\tilde x_{s})=\left\lVert x_{s} - \mathcal{S}_{\mathcal{D}_{s}}(\mathcal{S}_{\mathcal{E}_{s}}(x_{s}))\right\rVert$.

 \begin{figure*}
	\centering
    \includegraphics[scale=0.5]{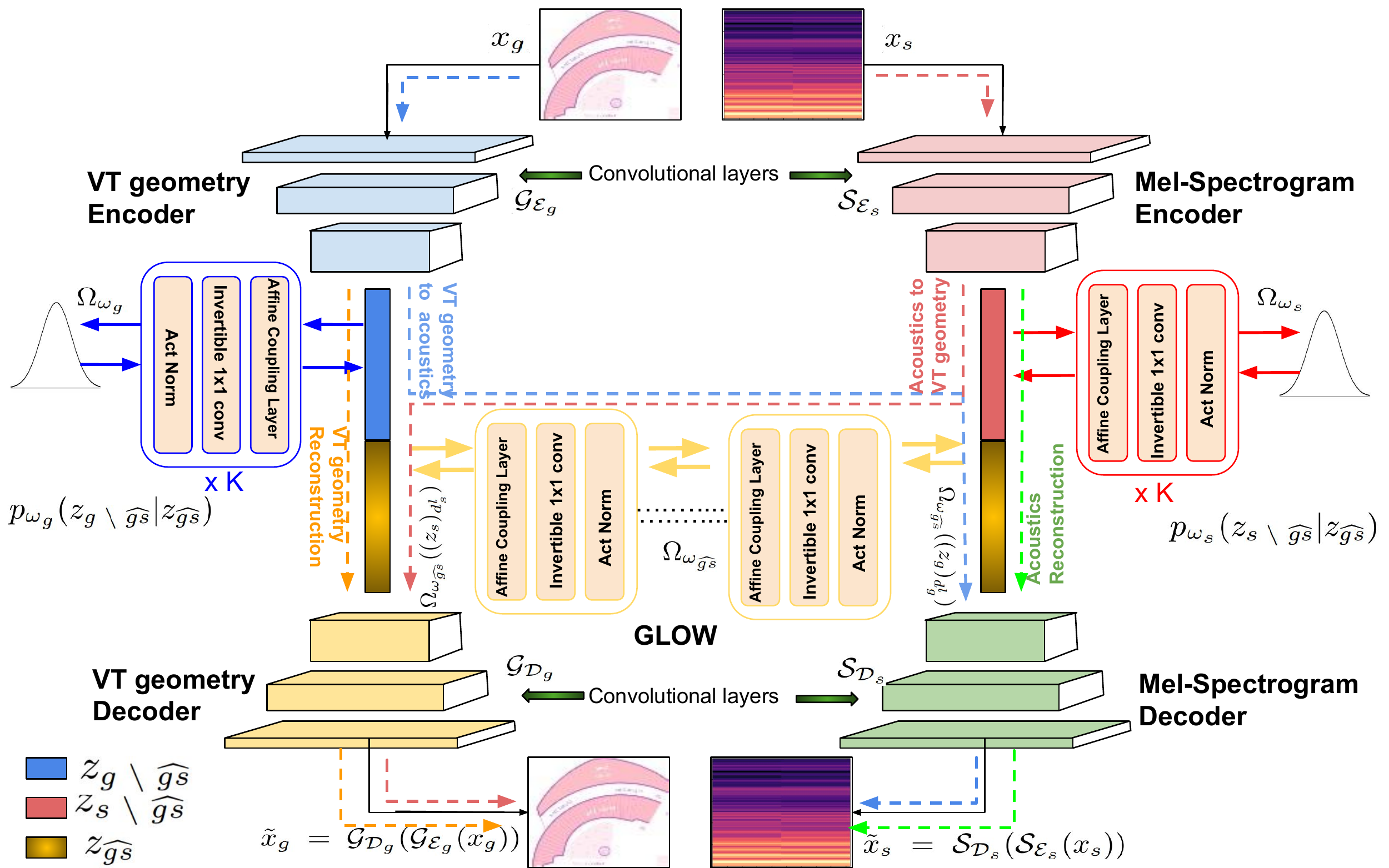}
    \caption{The proposed articulatory-acoustic forward and inverse mapping}

    \vspace{-7pt}
\end{figure*}

\subsection{Derivation of ELBO}
Since our entire latent variable representation is partitioned to three major components as discussed in section 2.1, accordingly, our posterior distribution gets factorized as follows: 
\begin{multline}
\Psi_{\phi}(z|x_{g},x_{s})=\Psi_{\gamma}(z_{g ~\setminus~ \widehat{gs}}|x_{g},z_{\widehat{gs}})\Psi_{\alpha}(z_{s ~\setminus~ \widehat{gs}}|x_{s},z_{ \widehat{gs}})\\\Psi_{\beta}(z_{\widehat{gs}}|x_{g},x_{s})
\end{multline}
Similarly, assuming the conditional independence of the latent codes, our prior probability distribution gets factorized as:
\begin{multline}
p(z)=p(z_{\widehat{gs}},z_{g~\setminus~ \widehat{gs}},z_{s~\setminus~ \widehat{gs}})= p(z_{g~\setminus~ \widehat{gs}}|z_{\widehat{gs}})~p(z_{s~\setminus~ \widehat{gs}}|z_{\widehat{gs}})\\~p(z_{\widehat{gs}})
\end{multline}

The computation of the optimal likelihood requires the marginalization of the latent variable, which is potentially challenging to compute. Instead, we optimize the lower bound over the encoding distribution using the standard procedure, as follows: 
\begin{multline}
\log p~(x_{g},x_{s}) = \log \left(\sum \Psi_{\phi}(z|x_{g},x_{s}) \frac{p~(x_{g},x_{s},z)}{\Psi_{\phi}(z|x_{g},x_{s})}\right)
\end{multline}
Now using Jensen's inequality, 
\begin{multline}
\log p~(x_{g},x_{s}) \geqslant \sum \Psi_{\phi}(z|x_{g},x_{s}) \log \frac{p~(x_{g},x_{s},z)}{\Psi_{\phi}(z|x_{g},x_{s})} \\= \sum \Psi_{\phi}(z|x_{g},x_{s}) \log \frac{p~(x_{g},x_{s}|z)p(z)}{\Psi_{\phi}(z|x_{g},x_{s})}\\\geqslant \sum \Psi_{\phi}(z|x_{g},x_{s}) [\log p~(x_{g},x_{s}|z) + \log p_{}(z)\\ - \log \Psi_{\phi}(z|x_{g},x_{s})]
\end{multline}

With the help of Equation (8) and (9), the first term or data-likelihood term in Equation (11), can be further simplified as:
\begin{multline}
\Psi_{\phi}(z|x_{g},x_{s}) \log p~(x_{g},x_{s}|z)=\Psi_{\gamma}(z_{g ~\setminus~ \widehat{gs}}|x_{g},z_{\widehat{gs}})\times\\\Psi_{\beta}(z_{\widehat{gs}}|x_{g},x_{s}) \log p(x_{g}|z_{\widehat{gs}},z_{g ~\setminus~ \widehat{gs}})~+\Psi_{\alpha}(z_{s ~\setminus~ \widehat{gs}}|x_{s},z_{\widehat{gs}})\times \\\Psi_{\beta}(z_{\widehat{gs}}|x_{g},x_{s}) \log p(x_{s}|z_{\widehat{gs}},z_{s ~\setminus~ \widehat{gs}})
\end{multline}
Similarly, the second term can be further simplified as:
\begin{multline}
 \Psi_{\phi}(z|x_{g},x_{s})\log p_{}(z)=\Psi_{\beta}(z_{\widehat{gs}}|x_{g},x_{s}) \log p(z_{\widehat{gs}})+\\\Psi_{\gamma}(z_{g ~\setminus~ \widehat{gs}}|x_{g},z_{\widehat{gs}}) \log p (z_{g ~\setminus~ \widehat{gs}}|z_{\widehat{gs}})+\Psi_{\alpha}(z_{s ~\setminus~ \widehat{gs}}|x_{s},z_{\widehat{gs}}) \times \\\log p (z_{s ~\setminus~ \widehat{gs}}|z_{\widehat{gs}}). 
 \end{multline}
 And the third term in equation (11) can be simplified as:
 \begin{multline}
 -\Psi_{\phi}(z|x_{g},x_{s})\log \Psi_{\phi}(z|x_{g},x_{s})=-\Psi_{\beta}(z_{\widehat{gs}}|x_{g},x_{s}) \times \\\log \Psi_{\beta}(z_{\widehat{gs}}|x_{g},x_{s})-\Psi_{\gamma}(z_{g ~\setminus~ \widehat{gs}}|x_{g},z_{\widehat{gs}}) \log \Psi_{\gamma}(z_{g ~\setminus~ \widehat{gs}}|x_{g},z_{\widehat{gs}})\\ - \Psi_{\alpha}(z_{s ~\setminus~ \widehat{gs}}|x_{s},z_{\widehat{gs}})\log \Psi_{\alpha}(z_{s ~\setminus~ \widehat{gs}}|x_{s},z_{\widehat{gs}}). 
  \end{multline}
 Therefore our task is now to maximize the lower bound \cite{kingma2013auto,tzikas2008variational,huang2020augmented, mahajan2020latent} obtained by plugging the expressions of Equations (12), (13) and (14) in Equation (11). 

\subsection{Normalizing flow}
Normalizing flow  \cite{huang2020augmented, rezende2015variational,dinh2016density,dinh2014nice,kingma2018glow,mahajan2020latent} is a flow-based generative model and is used as a powerful probability density estimator. It is constructed by stacking a sequence of invertible transformation functions which transform a simple distribution into a complex one and eventually learns an explicit data distribution $p(x)$. The probability distribution of the final target variable is obtained by substituting the variables for a new one, flowing through a chain of transformations $f_{i}$, following the change of variables theorem. Given an initial distribution $z_{0}$, the output $x$ can be obtained by using a series of probability density functions in a step-by-step fashion.
\begin{multline}
x=z_{K}=f_{K}(f_{K-1}(f_{K-2}(f_{K-3}(....(f_{3}(f_{2}(f_{1}(z_{0}))))....))))
\end{multline}
Using the change of variables rule, the probability density function of the model can therefore be written as follows:
\begin{multline}
\log ~p(x)=\log \pi_{K}(z_{K})=\log ~\pi_{K-1}(z_{K-1}) - \log  \left|{det \frac{df_{K}}{dz_{K-1}}}\right|\\=\log ~\pi_{0}(z_{0}) - \sum_{i=1}^{K} \log  \left|{det \frac{df_{i}}{dz_{i-1}}}\right|
\end{multline}
The sequence formed by successive distributions $\pi_{i}$ is known as normalized flow. Both the conditional priors $\Omega_{\omega_{g}}=\Psi_{\gamma}(z_{g ~\setminus~ \widehat{gs}}|x_{g},z_{\widehat{gs}})$ and $\Omega_{\omega_{s}}=\Psi_{\alpha}(z_{s ~\setminus~ \widehat{gs}}|x_{s},z_{\widehat{gs}})$ as well as the mapping between the shared latent codes $\Omega_{\omega_{\widehat{gs}}}$  are modeled with GLOW \cite{kingma2018glow}, a normalizing flow based generative model using invertible $1 \times 1$ convolutions. A single step of GLOW involves three substeps - activation normalization (act-norm), invertible $1 \times 1$ convolution and an affine coupling layer. The act-norm is an affine transformation using trainable parameters - scale $(s)$ and bias $(b)$ per channel, similar to batch normalization, except that it works for a mini-batch size 1. The transformation for a $k^{th}$ layer can be expressed as $y_{(k)~i,j}=s\odot z_{(k)~i,j} + b_{(k)}$. Next,  $1 \times 1$ convolution with equal input and output dimensions is a generalized way of permuting channel ordering between layers of flow, thereby ensuring that the ordering of channels is shuffled for the flow to act on the entire data sample. Assuming the weight matrix to be $W \colon [c \times c]$, where $c$ is the number of channels, this step can be written as $v_{(k)~i,j}=Wy_{(k)~i,j}$. The last substep consists of an affine coupling layer where the convolved outputs $v_{(k)}$ are split into two parts: $v_{(k)~a}$ and $v_{(k)~b}$, out of which $(v_{(k)~a})$ remains the same where as the other part $(v_{(k)~b})$ undergoes an affine transformation involving scaling $(s(.))$ and translation $(t(.))$. This can be denoted as: $v_{(k)~a},v_{(k)~b}= split (v_{(k)})$, $(log~s,t)=NN(v_{(k)~b})$, $u_{(k)~a}=v_{(k)~a}, ~u_{(k)~b}=exp(log~s)\odot v_{(k)~b}+t$, $z_{(k+1)}=concat(u_{(k)~a},u_{(k)~b})$. 
 \begin{figure}
	\centering
    \includegraphics[scale=0.25]{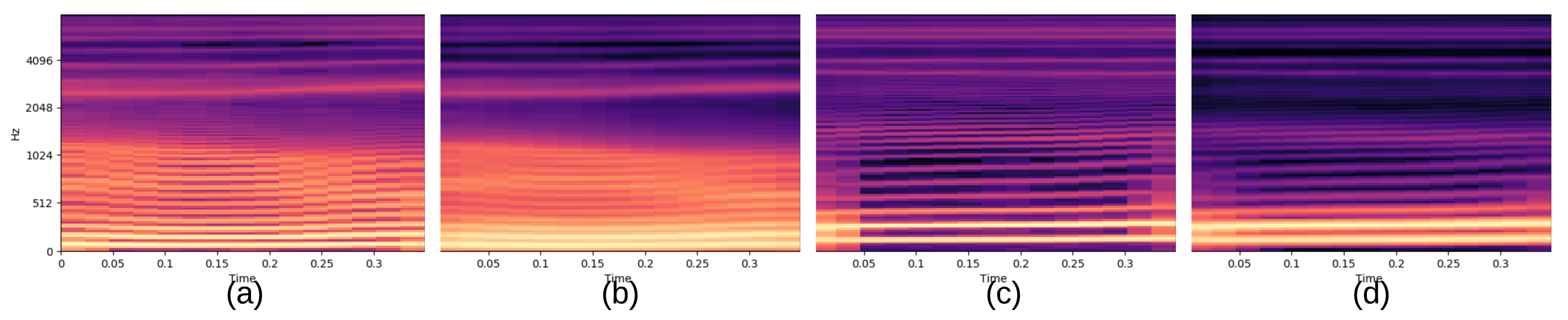}
    \caption{(a) and (c) respectively shows the mel-spectrogram corresponding to the original vowels /a/ and /u/, (b) and (d) respectively shows their synthesized versions from VT geometry }

    \vspace{-7pt}
\end{figure}

As shown in Fig 2, the mapping between the shared latent components $\Omega_{\omega_{\widehat{gs}}}$ is achieved using a sequence of such transformations. The cost of mapping the latent space of VT geometry image to that of mel-spectrogram $\mathcal{L}_{g2s}(x_{g},x_{s})$ is defined as mean squared error between the encoded spectrogram representation $(z_{s})_{d^{l}_{s}}$ and transformed image representation $\Omega_{\omega_{\widehat{gs}}}((z_{g})_{d^{l}_{g}})$. Similarly, the cost of mapping the latent space of mel-spectrogram to that of VT geometry image $\mathcal{L}_{s2g}(x_{s},x_{g})$ is defined as mean squared error between the encoded VT geometry representation $(z_{g})_{d^{l}_{g}}$ and transformed spectrogram representation $\Omega_{\omega_{\widehat{gs}}}((z_{s})_{d^{l}_{s}})$.


\section{Experiments and Results}
\subsection{Dataset and Training}
We varied the pink trombone tongue controller\footnote{https://dood.al/pinktrombone/} that changes the VT shape and correspondingly captured videos of pink trombone VT with frame rate of 30 fps and audios at a sampling rate of 22,020 Hz. Our model was implemented in PyTorch and we converted the audio into mel-spectrograms using Librosa \cite{mcfee2015librosa}. We randomly shuffled and partitioned the data (36,081 audios and images  extracted from the videos) into train (80\%), development (10\%) and test sets (10\%). The images were downsampled to dimensions $90\times 98 \times 3$ to reduce the computational time. The network was trained with a batch size of 10 on NVIDIA GeForce GTX 1080 Ti GPU. The loss function was optimized using Adam with a learning rate of .0001 for a total of 200 epochs. In order to mitigate the problem of overfitting, Batch Normalization was used after every convolutional layer and before applying non-linearity. 

 \begin{figure}
	\centering
    \includegraphics[scale=0.53]{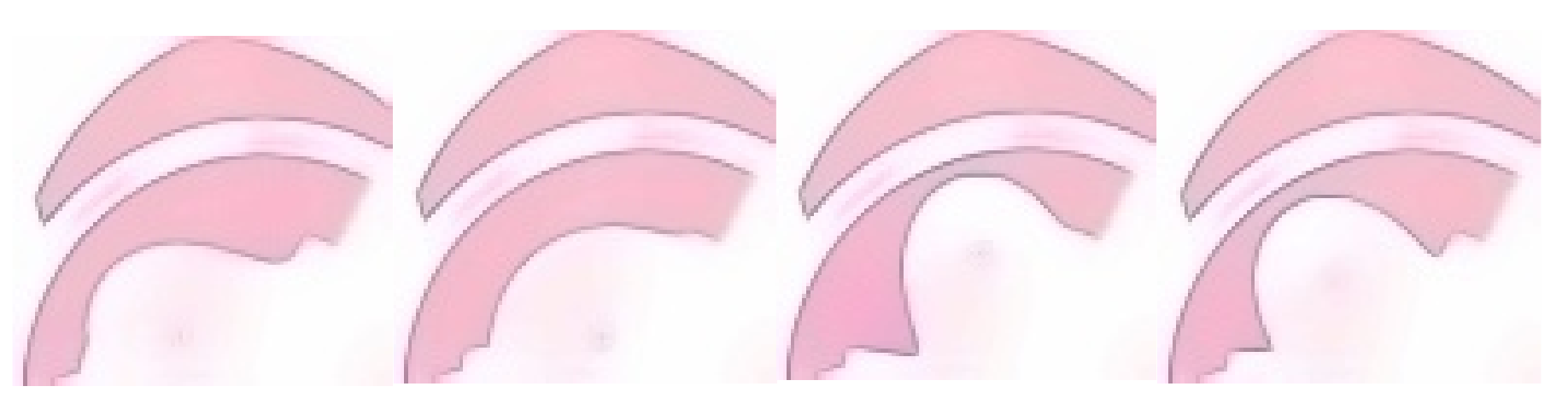}
    \caption{The synthesized pink trombone images corresponding to VT configurations for /a/, /ae/, /i/ and /u/ (left to right)}

    \vspace{-7pt}
\end{figure}

\subsection{Qualitative and quantitative performance analysis}
The original and synthesized Mel-Spectrograms of the vowels /a/ and /u/ corresponding to the respective VT shapes have been shown in Fig 3. A qualitative analysis of the figure demonstrates that although the generated mel-spectrograms are blurrier than the original crisp mel-spectrograms, they are indeed recognizable and significantly similar to the ground truth data. In order to quantitatively evaluate the performance of the proposed method in acoustic domain, we further computed the average formant frequencies of the synthesized audio signal and the original audio signal after recovering the synthesized audio with Griffin-Lim based spectrogram inversion method \cite{griffin1984signal}. The mean error of the first three formants of synthesized vowels w.r.t the original vowels are $18.57\%, 24.21\%, 7.69\%$ respectively.  The synthesized pink trombone images corresponding to the cardinal vowels $/a/$, $/ae/$, $/i/$ and $/u/$ have been presented in Fig. 4. The generated images are found to be quite similar to the actual VT geometries of pink trombone corresponding to respective vowels. It shows that our model is able to properly recognize the VT shape changes with changes in acoustic input, thereby demonstrating the success of our approach. The mean absolute error between the normalized synthesized pink trombone images and original images is $0.0397$, most of which evidently comes from the non-VT part.
\section{Conclusions and future works}
In this paper, we have developed a one-to-one invertible mapping between the articulatory and acoustic spaces for an online articulatory speech synthesizer application named pink trombone. To the best of our knowledge, this is the first attempt to study an invertible joint articulatory-acoustic representation utilizing the best of deep autoencoder architectures and normalizing flow based techniques. This can be extended to one-to-many or many-to-one scenarios by introducing variational autoencoder architecture which generates a vector of means and standard deviations of the Gaussian distributions as the latent codes and will be addressed in future works. Besides, this work investigates a joint articulatory-acoustic representation for static vowels only, as we are considering VT input image for a particular instant. This can be further extended to continuous vowel spaces by including a sequence of images reflecting the dynamic VT shape changes with time in the articulatory space. 

\section{Acknowledgements}
This work was funded by the Natural Sciences and Engineering Research Council (NSERC) of Canada and Canadian Institutes for Health Research (CIHR).

\bibliographystyle{IEEEtran}

\bibliography{aa}


\end{document}